\documentclass[12pt]{article}                          %
\usepackage{amsmath}                                   %
\usepackage{amsthm}                                    %
\usepackage{amssymb}                                   %
\usepackage{latexsym}                                  %
\usepackage{subeqnarray}
\newcommand{\nc}{\newcommand*}
\newcommand{\rnc}{\renewcommand*}
\theoremstyle{plain}
\newtheorem{thm}{Тheorem}
\def\cA{{\cal A}}
\def\cH{{\cal H}}
\def\cF{{\cal F}}
\def\cN{{\cal N}}
\def\one{1\hskip-.35em 1}
\rnc{\Bbb}{\mathbb}
\nc{\BR}{{\Bbb R}}
\nc{\BI}{{\Bbb I}}
\nc{\BC}{{\Bbb C}}
\nc{\reff}[1]{(\ref{#1})}% Put parentheses around equation references
\nc{\ts}{\textstyle}
\nc{\ds}{\displaystyle}
\nc{\ket}[1]{{\vert{#1}\rangle}}                % ket-vector
\nc{\bra}[1]{{\langle{#1}\vert}}                % bra-vector
\nc{\braket}[2]{{\langle{#1}\vert{#2}\rangle}}  % <x|y>
\nc{\ccr}[2]{\left[{#1}\,,\,{#2}\right]}        % commutator
\nc{\cel}[1]{[\hskip-.3em [\, #1 \,]\hskip-.3em ]}
\nc{\bcel}[1]{[\hskip-.1em [\, #1 \,]\hskip-.1em ]}
\nc{\Syx}[3]{\sum_{ #1 = #2 }^{#3}}
\nc{\Szx}[2]{\sum_{ #1 = 0 }^{#2}}
\nc{\Szi}[1]{\sum_{ #1 = 0 }^{\infty}}
\def\wt{\widetilde}
\def\wh{\widehat}
\def\half{{\frac{1}{2}}}
\def\ta{\wt{a}}
\def\tF{\wt{F}}
\def\tH{\wt{H}}
\def\tK{\wt{K}}
\def\tP{\wt{P}}
\def\tX{\wt{X}}
\def\HS{\widetilde{\cH}_{p,N}}
\def\hHS{\widehat{\cH}_{p,N}}
\def\HaS{{\cH}_{AS}}
\def\hmu{$\cH_{\mu}$}
\def\hmuN{$\cH_{\mu}^{(N)}$}
\def\lto{$\text{L}^2(\BR;\mu)$}
\def\intii{\int_{-\infty}^{\infty}}
\def\Kn{K_n(x;p,N)}
\def\tKn{{\tK}_n(x;p,N)}
\def\hKn{{\widehat{K}}_n(x;p,N)}
\nc{\tKx}[1]{{\tK}_{#1}(x;p,N)}
\nc{\tKy}[1]{{\tK}_{#1}(y;p,N)}
\nc{\tghfto}[3]{{{}_{2}{\tF}_{1}}\left(\ts{\genfrac{}{}{0pt}{}{#1}{#2}\biggl.\biggr| #3 }\right)}
\begin{document}

{}
\bigskip
\centerline{\large\bf Vadim V. Borzov
%${}^{(1)}$
\normalsize{and}
\large\bf Eugene V. Damaskinsky
%${}^{(2)}$
}

\bigskip

\centerline{\Large\bf Coherent States for}
\bigskip

\centerline{\Large\bf  Generalized oscillator with}
\bigskip

\centerline{\Large\bf Finite-Dimensional Hilbert Space
\footnote{This research was partially supported by RFBR grant
No 06-01-00451}}

\bigskip

\begin{abstract}
The construction of oscillator-like systems connected with the
given set of orthogonal polynomials and coherent states for
such systems developed by authors is extended to the case of
the systems with finite-dimensional state space. As example we
consider the generalized oscillator connected with Krawtchouk
polynomials.

%This investigation is partially supported by RFFI grants No 06-01-00451.
\end{abstract}
\bigskip

\section {Introduction}
In the last years the interest to studying of the oscillator-like
systems (so-called "generalized oscillators") and to using such
systems in various areas of quantum mechanics (\cite{1}-\cite{3})
has significantly increased. In works \cite{4}-\cite{11} authors
suggested a new approach to definition of generalized oscillators,
connected with the given system of orthogonal polynomials and to
construction of coherent states for these oscillators. Within the
framework of this approach we investigated generalized oscillators,
connected with classical orthogonal polynomials of a continuous
argument (such as Laguerre, Legendre, Chebyshev, and Gegenbauer
polynomials), orthogonal polynomials of a discrete argument
(such as Meixner and Charlier polynomials) as well as systems
connected with $q$-analogues of Hermite polynomials. All these
oscillator-like systems defined in an infinite-dimensional state
space. However many interesting applications in nonlinear quantum
optics are connected with finite-dimensional state spaces~\cite{12}.
In these considerations the Krawtchouk polynomials, which form the
important class of classical orthogonal polynomials of discrete
argument are frequently used. Convenience of these polynomials
for applications is connected with the fact that these polynomials
can be considered as a finite-dimensional approximation of the
Hermite and Charlier polynomials~\cite{13}. The generalized
oscillator connected with Krawtchouk polynomials~\cite{14}-\cite{16}
is a typical example of generalized oscillator in finite-dimensional
Hilbert space.

Within the framework of our approach, we discuss here the construction
of generalized oscillator in finite-dimensional Fock space and
definition of its coherent states in a such way that in the limiting
case when dimension of space goes to infinity these coherent states
become ones of appropriate generalized oscillator with infinite state
space\footnote{In the literature the oscillator with finite-dimensional
state space is usually named - finite oscillator and we frequently called
them FD-oscillators for brevity.}. In the finite-dimensional state
space the annihilation operator has only one eigenvector with zero
eigenvalue (the vacuum state), so the standard definition of
Barut-Girardello coherent states as eigenstates of the annihilation
operator is inapplicable. There are several variants of definition of
coherent states for finite-dimensional analogue of the usual boson
oscillator (the spin coherent states, phase coherent states etc.\cite{12}).
In present work we consider a generalization of the definition given
in~\cite{17},\cite{18} for the standard (boson like) FD-oscillator.
This generalization is a modification of the Glauber definition \cite{Gl}
to the case of "truncated" exponential operators. The coherent states
obtained by such construction can be considered as  coherent states
of Klauder - Gazeau type \cite{KG}.

\underline{\textbf 1.} We recall, following \cite{4}, the basic
steps of our construction of oscillator-like systems. For brevity we
discuss here only a symmetric case, when the Jacobi matrix defined by
recurrent relations for considered orthogonal polynomials has a zero
diagonal (in this case an orthogonality measure is symmetric under
reflection with respect to the origin of coordinates).

Let us denote by \hmu the Hilbert space \lto, where $\mu$ is a positive
symmetric probability Borel measure on the real axis. We suppose that
all power moments of this measure are finite and
\begin{equation}\label{1}
\mu_0=\intii\mu(\rm{d}x)=1,\qquad
\mu_{2k+1}=\intii x^{2k+1}\mu(\rm{d}x)=0.
\quad k=0,1,\ldots .
\end{equation}
We define a positive numerical sequence
$$
\left\{b_{n}\right\}_{n=0}^{\infty}, \quad
b_{n}>0, \quad b_{-1}=0
$$
as the solution of the algebraic system
\begin{equation}\label{2}
\Szx{m}{\cel{\frac{n}{2}}}\Szx{s}{\cel{\frac{n}{2}}}
\frac{(-1)^{n+s}}{\left(b_{n-1}^{\,\,2}\right)!}
\alpha_{2m-1, n-1}\alpha_{2s-1, n-1}\mu_{2(n-m-s+1)}=
b_{n-1}^{\,\,2}+b_{n}^{\,\,2},\quad n\geq0,
\end{equation}
where
\begin{gather}
\mu_{2k}=\intii x^{2k}\mu(\rm{d}x),\quad k=0,1,\ldots ,
\label{3}\\
\alpha_{-1, -1}=0,\qquad\alpha_{2p-1, n-1}=
\Syx{k_1}{2p-1}{n-1}b_{k_1}^{\,\,2}
\Syx{k_2}{2p-3}{k_1-2}b_{k_2}^{\,\,2}\cdot\ldots\cdot
\Syx{k_p}{1}{k_{p-1}-2}b_{k_p}^{\,\,2} .\label{4}
\end{gather}
Here the symbol $\bcel{x}$ - denotes the whole part of $x,$
and "a factorial on an index"
$
\left(b_{n-1}^{\,\,2}\right)!
$
is defined by relation
$
\left(b_{n-1}^{\,\,2}\right)!=
b_{0}^{\,\,2}b_{1}^{\,\,2}\cdot\ldots\cdot b_{n-1}^{\,\,2}.
$

It is easy to check, that the relations
\begin{equation}\label{5}
b_{0}^{\,\,2}=\mu_2,\quad b_{1}^{\,\,2}=\frac{\mu_4}{\mu_2}-\mu_2,
\,\ldots ,
\end{equation}
give the unique solution of the system \reff{2}.

The system of polynomials
$
\left\{\psi_n(x)\right\}_{n=0}^{\infty}
$
is called a canonical system associated with the measure $\mu,$ if
the following recurrent relations are fulfilled
\begin{gather}
x\psi_{n}(x)=b_{n}\psi_{n+1}(x)+b_{n-1}\psi_{n-1}(x),\quad n\geq0,
\quad b_{-1}=0 , \label{6}\\
\psi_{0}(x)=1, \label{7}
\end{gather}
where the positive coefficients
$
\left\{b_n\right\}_{n=0}^{\infty}
$
are solutions of the system \reff{2}.

\begin{thm} [see \cite{4}] Let the system of polynomials
$
\left\{ {\psi_{n}(x)}\right\}_{n=0}^{\infty }
$
satisfies to the relations \reff{6} and \reff{7} with coefficients
$
\left\{ {b_{n}}\right\}_{n=0}^{\infty } ,
$
forming a positive sequence and let $\mu$ be
some symmetric probability measure on $\BR.$ This system of
polynomials is orthogonal with respect to the measure $\mu$
if and only if the polynomial system
$
\left\{ {\psi_{n}(x)}\right\}_{n=0}^{\infty }
$
is the canonical polynomial system associated with the measure $\mu$
(i.e. the coefficients
$
\left\{ {b_{n}}\right\}_{n=0}^{\infty } ,
$
are solutions of the system \rm{(\ref{2})}).
\end {thm}

\medskip

We define the selfadjointed operators: "coordinate"
$X_{\mu},$ "momentum" $P_{\mu}$ and quadratic Hamiltonian
$
\rm{H}_{\mu}=X_{\mu}^{\, 2}+P_{\mu}^{\, 2}
$
by their action on elements of the basis
$
\left\{ {\psi_{n}(x)}\right\}_{n=0}^{\infty }
$
in the Hilbert space \hmu, according to formulas
\begin{align}
X_{\mu}\psi_{0}(x)&=b_{0}\psi_{1}(x),\quad
X_{\mu}\psi_{n}(x)=b_{n}\psi_{n+1}(x)+b_{n-1}\psi_{n-1}(x),\quad n\geq 1,
\label{8}\\
P_{\mu}\psi_{0}(x)&=\!-ib_{0}\psi_{1}(x),\quad\!\!
P_{\mu}\psi_{n}(x)\!=i\left(b_{n-1}\psi_{n-1}(x)-b_{n}\psi_{n+1}(x)\right),
\,\, n\geq\! 1,\label{9}\\
\rm{H}_{\mu}\psi_{n}(x)&=\lambda_n\psi_{n}(x),\qquad n\geq 0,
\label{10}
\end{align}
where
\begin{equation}\label{11}
\lambda_{0}=2{b_{0}^{2}},\qquad
\lambda_{n}=2({b_{n-1}^{2}}+{b_{n}^{2}}),\qquad n\geq 1.
\end{equation}
We define further the creation and annihilation operators
by the standard relations
\begin{equation}\label{B11}
a^{+}_{\mu}:=\frac{1}{\sqrt{2}}\left(X_{\mu}+iP_{\mu}\right),\qquad
a^{-}_{\mu}:=\frac{1}{\sqrt{2}}\left(X_{\mu}-iP_{\mu}\right),
\end{equation}
These operators act on the elements of the basis in Hilbert space
$\cH,$ according to the relations
\begin{equation}\label{B12}
a^{+}_{\mu}\psi_{n}(x)=\sqrt{2}b_{n}\psi_{n+1}(x),\qquad
a^{-}_{\mu}\psi_{n}(x)=\sqrt{2}b_{n-1}\psi_{n-1}(x),\quad
n\geq 0,\quad b_{-1}=0.
\end{equation}

We shall consider Hilbert space \hmu as a functional realization
of the Fock space  $\cF$ with basis
$
\left\{\ket{n}\right\}_{n=0}^{\infty},
$
so that
$
\left\{ {\psi_{n}(x)}\right\}_{n=0}^{\infty }
$
(where $\psi_{n}(x)=\braket{x}{n}$) gives realization of this
basis in "coordinate" representation. We define also the
self-ajointed operator $\cN,$ "numbering" basic elements,
\begin{equation}\label{14}
\cN\ket{n}=n\ket{n},\quad n\geq 0,
\end{equation}
and the operator $B(\cN),$ acting on basic vectors,
according to formulas
\begin{equation}\label{15}
B(\cN)\ket{n}=b_{n-1}^{\,\,2}\ket{n},\quad n\geq 0,
\end{equation}
so that
\begin{equation}\label{16}
B(\cN+I)\ket{n}=b_{n}^{\,\,2}\ket{n},\quad n\geq 0,
\end{equation}
Let us note that it is not supposed, that
$\cN=a^{+}_{\mu}a^{-}_{\mu}.$

It is simple to check the validity of commutation relations
\begin{equation}\label{17}
\left[a^{-}_{\mu},a^{+}_{\mu}\right]=2\left(B(\cN+I)-B(\cN)\right),\qquad
\left[\cN,a^{\pm}_{\mu}\right]=\pm a^{\pm}_{\mu} .
\end{equation}
We shall call an algebra
$\cA_{\mu},$ generated by operators
$
a^{\pm}_{\mu},\,\cN
$
with commutation relations \reff{17},
the generalized oscillator connected with given system of polynomials
$
\left\{ {\psi_{n}(x)}\right\}_{n=0}^{\infty}
$
orthonormalized with respect to the symmetric probability measure $\mu$.
Below we shall use the symbol $\cA_{\mu}$ to denote the
generalized oscillator  as well.
\bigskip

\noindent\underline{\textbf 2.} In a general case when the measure
$\mu$ is not symmetric, it is possible to find  from the given sequence
of its moments
$
\left\{ {\psi_{n}(x)}\right\}_{n=0}^{\infty},
$
($\mu_{0}=1$) two uniquely defined sequences of real numbers
$
\left\{ a_{n}\right\}_{n=0}^{\infty}
$
and
$
\left\{ b_{n}\right\}_{n=0}^{\infty}
$
and define a canonical system of polynomials
$\left\{ {\psi_{n}(x)}\right\}_{n=0}^{\infty},$
which are orthogonal
with respect to a measure $\mu$ and satisfy recurrent relations
\begin{equation}\label{18}
x\psi_{n}(x)=b_{n}\psi_{n+1}(x)+a_{n}\psi_{n}(x)+b_{n-1}\psi_{n-1}(x),
\qquad  n\geq 0,\quad b_{-1}=0,
\end{equation}
with the initial condition
\begin{equation}\label{19}
\psi_{0}(x)=1 .
\end{equation}

Together with this system
$
\left\{\psi_{n}(x)\right\}_{n=0}^{\infty}
$
of polynomials, orthogonal with respect to a measure $\mu,$ we shall
consider one more system of polynomials
$
\left\{\psi_{n}^{(0)}(x)\right\}_{n=0}^{\infty},
$
which are orthogonal with respect to another symmetric measure $\mu^0$
and satisfy recurrent relations
\begin{equation}\label{20}
x\psi_{n}^{(0)}(x)=b_{n}\psi_{n+1}^{(0)}(x)+b_{n-1}\psi_{n-1}^{(0)}(x),
\qquad  n\geq 0,\quad b_{-1}=0,
\end{equation}
with the initial condition
\begin{equation}\label{21}
\psi_{0}(x)=1 .
\end{equation}
Let us denote by $\cA_{\mu^0}$ the generalized oscillator, constructed
by the method described above in the section \underline{\textbf 1.} and
defined in the space
$
\cF_{\mu^0}\left(=\cH_{\mu^0}\right).
$

To construct an oscillator $\cA_{\mu},$ connected with the initial
measure $\mu,$ it is necessary to introduce, following \cite{4},
the generalized coordinate operator $\wt{X}_{\mu},$
the generalized momentum operator $\wt{P}_{\mu}$ and Hamiltonian
$\wt{H}_{\mu}$ as well as ladder creation and annihilation operators
$\wt{a}_{\mu}^{\pm},$ according to formulas
\begin{gather}
\wt{X}_{\mu}=\rm{Re}\left(X_{\mu}-P_{\mu}\right),\quad
\wt{P}_{\mu}=-i\rm{Im}\left(X_{\mu}-P_{\mu}\right),\quad
\wt{H}_{\mu}=\wt{X}_{\mu}^{\,\,2}+ \wt{P}_{\mu}^{\,\,2},\label{22}\\
\wt{a}_{\mu}^{\pm}=
\frac{1}{\sqrt{2}}\left(\wt{X}_{\mu}\pm\wt{P}_{\mu}\right),\label{23}
\end{gather}
and to define the operator $\wt{\cN}_{\mu}.$
Operators $X_{\mu}$ and $P_{\mu}$ are defined by relations
\reff{8} and \reff{9}, but using the commutation relations
\reff{18} instead of \reff{6} (see \cite{4}).

\bigskip

It is follows from the results of work \cite{4} that an
action of operators
$
\wt{a}_{\mu}^{\pm}, \wt{\cN}_{\mu}
$
on the basic elements of Fock space
$\cF_{\mu}$ can be defined by the same relations as an
action of operators
$
\wt{a}_{\mu^0}^{\pm}, \wt{\cN}_{\mu^0}
$
in the space $\cF_{\mu^0},$
so operators
$
\wt{a}_{\mu}^{\pm}, \wt{\cN}_{\mu}
$
satisfy in space $\cF_{\mu}$ the same commutation relations \reff{17}
as operators
$
\wt{a}_{\mu^0}^{\pm}, \wt{\cN}_{\mu^0}
$
in $\cF_{\mu^0}.$
Hence, $\cA_{\mu}$ and $\cA_{\mu^0}$
define unitary equivalent representations of the same
generalized oscillator algebra.

\bigskip

\noindent\underline{\textbf 3.} In works \cite{12}, \cite{17}, \cite{18}
were considered oscillators, with finite-dimensional state spaces
(FD-oscillators) and which become the usual boson oscillator when
the dimension $N$ of this space goes to infinity. In the present work we
shall consider FD-oscillators which in the limit $N\rightarrow\infty$
become the generalized oscillator $\cA_{\mu}$  described above.
In consideration of FD-oscillators we shall use the following notations.
We denote by $\cH_{\mu}^{(N)}$ the $(N+1)$-dimensional subspace of
the Hilbert space
$
\cF_{\mu}\left(=\cH_{\mu}\right)
$
spanned by the first $N+1$ states
$\ket{0},\ket{1},\ldots,\ket{N}.$ In the space $\cH_{\mu}^{(N)}$
the following  orthogonality and completeness relations are fulfilled
\begin{equation}\label{24}
\braket{n}{m}=\delta_{n, m},\qquad
\Syx{n}{1}{N}\ket{n}\bra{n}=\one_{N+1}.
\end{equation}

The creation and annihilation operators in the space $\cH_{\mu}^{(N)}$
are defined by relations
\begin{align}
a_N^-:&=\Syx{k}{1}{N}\sqrt{2}b_{k-1}\ket{k-1}\bra{k},\label{25}\\
a_N^+:&=\Syx{k}{1}{N}\sqrt{2}b_{k-1}\ket{k}\bra{k-1}.\label{26}
\end{align}
They act on elements of basis in the natural way. Operators
$\cN_N$ and $B(\cN_N)$ are defined in the same way as above.
The commutation relations of operators $a_N^{\pm}, \cN_N$ in
space $\cH_{\mu}^{(N)}$ look like
\begin{equation}\label{27}
\begin{aligned}
\ccr{a_N^-}{a_N^+}\,&=2\left(B(\cN_N+\one_{N+1})-B(\cN_N)\right)-
2b_N^{\,\,2}\ket{N}\bra{N},\\
\ccr{\cN_N}{a_N^{\pm}}&=\pm a_N^{\pm}.
\end{aligned}
\end{equation}

In the following section we shall describe the Krawtchouk
oscillator as an important example of FD-oscillators.

\section{Krawtchouk oscillator}
\underline{\textbf 1. \, Krawtchouk polynomials }
The Krawtchouk polynomials can be defined with help of the
hypergeometric function (see \cite{13})
\begin{equation}\label{28}
\Kn:=\tghfto{-n,-x}{-N}{p^{-1}}=\Szx{k}{N}\frac{(-n)_k(-x)_k}{k!(-N)_kp^k}
\end{equation}
where
$
0<p<1,$\quad $n=0,1,\ldots,N,
$
and where the Pochhammer symbol $(a)_k $ is defined by the relation
$$
(a)_0=1,\quad (a)_k=a(a+1)\cdot\ldots\cdot(a+k-1)=
\frac{\Gamma(a+k)}{\Gamma(a)} .
$$

For definition of generalized oscillator it is convenient
(following \cite{4}) to redefine polynomials $\Kn$ as follows
\begin{equation}\label{29}
\tKn=\sqrt{\rho(n;p,N)}\Kn ,\quad n=0,1,\ldots,N ,
\end{equation}
where
\begin{equation}\label{30}
\rho(n;p,N)=C_N^{\xi}p^{\xi}(1-p)^{N-\xi} , \qquad
C_N^{\xi}:=\frac{N!}{\Gamma(\xi+1)\Gamma(N-\xi+1)} .
\end{equation}
The redefined polynomials satisfy to recurrent relations with
symmetric Jacobi matrix ($n=0,1,\ldots,N,$)
\begin{align}
x\tKn=b_n\tKx{n+1}&+a_n\tKn+b_n\tKx{n-1} , \label{31}\\
\tKx{0}&=1 ,  \label{32}
\end{align}
where
\begin{equation}\label{33}
a_n=p(N-n)+n(1-p),\qquad b_n=-\sqrt{p(1-p)(n+1)(N-n)},
\end{equation}
Together with polynomials $\tKn$ we shall use their modified variant
\begin{equation}\label{34}
\hKn=(-1)^n\tKn, ,\quad n=0,1,\ldots,N .
\end{equation}

The Krawtchouk polynomials $y(x)=\Kn$ satisfy the difference equation
\begin{equation}\label{35}
-ny(x)=p(N-x)y(x+1)-[p(N-x)+x(1-p)]y(x)+x(1-p)y(x-1) ,
\end{equation}
which remains valid also for polynomials $\tKn$ and $\hKn.$
The Krawtchouk polynomials fulfill the following
orthogonality relations
\begin{subequations}\label{36}
\begin{eqnarray}
\Szx{x}{N}\rho(x;p,N)\tKx{m}\tKn=\delta_{m,n}, \\
\Szx{n}{N}\rho(n;p,N)\tKx{n}\tKy{n}=\delta_{x,y}.
\end{eqnarray}
\end{subequations}
\bigskip

\underline{\textbf 2. The Krawtchouk oscillator. \,} Now we
define the Krawtchouk oscillator, according to general
construction of oscillator-like system discussed above.

Let us denote by
$
\HS=\ell^2_{N+1}(\rho(x;p,N))
$
the $N+1$-dimensional Hilbert space spanned by orthonormalized
(with respect to weight function $\rho(x;p,N)$) basis
$
\left\{\tKn\right\}_{n=0}^N
$
(by the symbol $\hHS$ we shall denote the same space in a case
when polynomials $\left\{\hKn\right\}_{n=0}^N$ are used
as orthonormalized basis in this space).

Let's call the Krawtchouk oscillator the oscillator-like system
determined by coordinate $\tX$ and momentum $\tP$ operators and
quadratic Hamiltonian $\tH$
\begin{equation}\label{37}
\tX:=\text{Re}(X-P),\quad
\tP:=-i\text{Im}(X-P),\quad
\tH:=\frac{1}{4p(1-p)}\left(\tX^{2}+\tP^{2}\right)
\end{equation}
where operators $X$ and $P$ act on elements of basis
$\left\{\tKn\right\}_{n=0}^N$
in the space $\HS$ according to formulas $(n\geq1)$
\begin{align}
\!\!\!\!\!\!X\tKx{0}&=b_0\tKx{1}+a_0\tKx{0}, \label{38} \\
\!\!\!\!\!\!P\tKx{0}&=-ib_0\tKx{1}+a_0\tKx{0}, \label{39} \\
\!\!\!\!\!\!X\tKx{n}&=\!b_n\tKx{n+1}\!+\!a_n\tKx{n}\!+\!b_{n-1}\tKx{n-1},
\label{40} \\
\!\!\!\!\!\!P\tKx{n}&=\!-ib_n\tKx{n+1}\!+\!a_n\tKx{n}\!+\!ib_{n-1}\tKx{n-1},
\label{41}
\end{align}
and coefficients $a_n$ and $b_n$ are determined by
formulas \reff{33}.
The  creation and annihilation operators
\begin{equation}\label{42}
\ta^{+}:=\frac{1}{2\sqrt{p(1-p)}}\left(\tX+i\tP\right),\qquad
\ta^{-}:=\frac{1}{2\sqrt{p(1-p)}}\left(\tX-i\tP\right)
\end{equation}
act on basic elements of space $\HS$ according to formulas
\begin{align}
\ta^{-}\tKx{n}&=-\sqrt{n(N-n+1)}\tKx{n-1},\label{43}\\
\ta^{+}\tKx{n}&=-\sqrt{(n+1)(N-n)}\tKx{n+1}.\label{44}
\end{align}
These operators satisfy the commutation relations
\begin{equation}\label{45}
\left[\ta_{K}^{-}, \ta_{K}^{+}\right]=\left(N-\BI-2\cN\right)
\end{equation}
where $\cN$ - the operator numbering basis elements
\begin{equation}\label{46}
\cN\tKx{n}=n\tKx{n} .
\end{equation}
Hamiltonian $\tH$ can be written in the form
\begin{equation}\label{47}
\tH=\half\left(\ta^{+}\ta^{-}+\ta^{-}\ta^{+}\right).
\end{equation}
From the results of work~\cite{4} it follows, that
this Hamiltonian has the spectrum
\begin{equation}\label{48}
\lambda_n=N(n+\half)-n^2 ,\quad  (0\leq n\leq N),
\end{equation}
so that
$$
\lambda_0=\half N=\lambda_N .
$$

Analogously it is possible to define modified oscillator in
the space $\hHS$ and check that these two oscillators are unitary
equivalent (the creation and annihilation operators and Hamiltonian
act identically on the appropriate bases).

Using results of work \cite{5} it is possible to show that the
eigenvalue equation
$
\tH y=\lambda_ny
$
in the space $\HS$ is equivalent to the difference equation \reff{35}.
To receive an exact form of the Hamiltonian $\tH$ in the space $\HS$
it is helpful to compare our definition of the Krawtchouk oscillator
with the variant considered in ~\cite{14} and ~\cite{15}.
\bigskip

\underline{\textbf 3. Krawtchouk oscillator defined in \cite{14}. \,}
In works ~\cite{14} and ~\cite{15} was considered another variant of
the Krawtchouk oscillator with Hamiltonian
\begin{equation}\label{49}
H_{AS}^{K}(\xi)=2p(1-p)N+\half +(1-2p)\frac{\xi}{h}-
\sqrt{p(1-p)}\left[
\alpha(\xi)e^{h\partial_{\xi}}+
\alpha(\xi-h)e^{-h\partial_{\xi}}
\right],
\end{equation}
where
\begin{equation}\label{50}
h=\sqrt{2Np(1-p)},\qquad
\alpha(\xi)=
\sqrt{\left(
(1-p)N-\frac{\xi}{h}\right)\left(pN+1+\frac{\xi}{h}
\right)} .
\end{equation}
The operator $H_{AS}$ is defined in the Hilbert space
$\HaS=\ell^2(\xi)$ with basis formed by Krawtchouk functions
\begin{equation}\label{51}
\Psi_{n}(\xi)=(-1)^n\sqrt{C_N^n\left(\frac{p}{1-p}\right)^n
\rho(pN+\frac{\xi}{h};p,N)}\,K_n(pN++1+\frac{\xi}{h});p,N) ,
\end{equation}
which satisfy two (dual) orthogonality relations
\begin{equation}\label{52}
\Szx{j}{N}\Psi_{n}(\xi_j)\Psi_{m}(\xi_j)=\delta_{n\,m},\qquad
\Szx{j}{N}\Psi_{j}(\xi_m)\Psi_{j}(\xi_n)=\delta_{n\,m},
\end{equation}
$\xi_j=h(j-pN),\quad j=0,1,\ldots,N.$
These Krawtchouk  functions are eigenfunctions for Hamiltonian $H_{AS}$
\begin{equation}\label{53}
H_{AS}(\xi)\Psi_{n}(\xi)=\lambda_n\Psi_{n}(\xi); \qquad
\lambda_n=n+\half,\quad n=0,1,\ldots,N ,
\end{equation}
which can be factorised
\begin{equation}\label{54}
H_{AS}^{K}(\xi)=\half\left[A^+,A^-\right]+\half (N+1) ,
\end{equation}
with the help of operators
\begin{align}
A^+(\xi)&=(1-p)e^{-h\partial_{\xi}}\alpha(\xi)-
p\alpha(\xi)e^{h\partial_{\xi}}+\nonumber\\
{}&\qquad\qquad
\sqrt{p(1-p)}\left((2p-1)N+\frac{2\xi}{h}\right), \label{55}\\
A^-(\xi)&=(1-p)\alpha(\xi)e^{h\partial_{\xi}}-
pe^{-h\partial_{\xi}}\alpha(\xi)+\nonumber\\
{}&\qquad\qquad
\sqrt{p(1-p)}\left((2p-1)N+\frac{2\xi}{h}\right). \label{56}
\end{align}
These operators act on the elements of basis
$\left\{\Psi_{n}(\xi)\right\}_{n=0}^{N}$
according to
\begin{equation}\label{57}
\begin{aligned}
A^+(\xi)\Psi_{n}^K(\xi)&=\sqrt{(n+1)(N-n)}\Psi_{n+1}^K(\xi) ,\\
A^-(\xi)\Psi_{n}^K(\xi)&=\sqrt{n(N-n+1)}\Psi_{n-1}^K(\xi) .
\end{aligned}
\end{equation}
Operators $A^+(\xi)$ and $A^-(\xi)$ together with the commutator
\begin{equation}\label{58}
A_0(\xi):=\half\left[A^+(\xi),\,A^-(\xi)\right]
\end{equation}
satisfy the commutation relations of the $so(3)$ algebra
\begin{equation}\label{59}
\left[A_0(\xi),\,A^{\pm}(\xi)\right]={\pm}A^{\pm}(\xi),\quad
\left[A^+(\xi),\,A^-(\xi)\right]=2A_0(\xi) .
\end{equation}

\bigskip
\underline{\textbf 4. Connection of two variants of
Krawtchouk oscillator. \,}
In addition to considered above Hilbert spaces
$
\HS=\ell^2_{N+1}(\rho(x;p,N)),
$
$
\widehat{\cH}_{p,N}
$
and
$
\HaS=\ell^2(\xi)
$
with bases
$
\left\{\tKn\right\}_{n=0}^N,
$
$
\left\{\hKn\right\}_{n=0}^N
$
and
$
\left\{\Psi_{n}(\xi)\right\}_{n=0}^{N},
$
respectively, we shall use auxiliary Hilbert space
$
\wh{\cH}_{AS}=\ell^2_{N+1}(\wt{\rho}(\xi))
$
with basis
$
\left\{\wh{K}_n(pN+\frac{\xi}{h};p,N)\right\}_{n=0}^N
$
orthogonal with respect to weight
$
\wt{\rho}(\xi)=\rho(pN+\frac{\xi}{h};p,N).
$

Further we define unitary operators $U,\,$ $V$ and $W,$
by the relations
\begin{align}
U\tKn&=\hKn,\quad U:\HS\rightarrow\widehat{\cH}_{p,N}^{K};
\label{60}\\
V\wh{K}_n(pN\!+\!\frac{\xi}{h};p,N)&\!=\!
\wt{\rho}(\xi)\wh{K}_n(pN\!+\!\frac{\xi}{h};p,N)\!=\!\Psi_{n}(\xi),\quad
V\!:\wh{H}_{AS}\!\rightarrow\!\HaS;
\label{61}\\
W\hKn&=\wh{K}_n(pN+\frac{\xi}{h};p,N),\quad
W:\widehat{\cH}_{p,N}\rightarrow\wh{\cH}_{AS}.
\label{62}
\end{align}
Then the similarity  transformation with the unitary operator
\begin{equation}\label{63}
T:=VWU,\qquad T:\HS\rightarrow\wh{\cH}_{AS}
\end{equation}
realize unitary equivalence of the operator $B_{AS}$
in $\HaS$ with the operator $\wt{B}$ in $\HS$
\begin{equation}\label{64}
\wt{B}=T^{-1}B_{AS}T.
\end{equation}

Using the operators $\ta^{\pm}$ \reff{42}
in space $\HS,$ we define operators
\begin{equation}\label{65}
\tK_{\pm}=\frac{1}{\sqrt{2p(p-1)}}\ta^{\pm},\qquad
\tK_{0}=\half\left[\tK_{-}\tK_{+}\right].
\end{equation}
These operators satisfy the commutation  relations of the $so(3)$
Lie algebra
\begin{equation}\label{66}
\left[\tK_{0},\,\tK_{\pm}\right]={\pm}\tK_{\pm},\quad
\left[\tK_{+},\,\tK_{-}\right]=2\tK_{0}
\end{equation}
and are unitary equivalent to the operators  $A^{\pm}(\xi)$
and $A_0(\xi)$
\begin{equation}\label{67}
A^{\pm}(\xi)=T\tK_{\pm}T^{-1},\qquad A_0(\xi)=T\tK_{0}T^{-1} .
\end{equation}

Using the relations \reff{67} and the explicit form
\reff{55}-\reff{56} of operators $A^{\pm}(\xi),$
it is possible to find an explicit
expressions for operators $\tK_{0}$ and $\tK_{\pm},$ and then
for the basic operators $\ta^ {\pm}$ and $\tH$ (and also operators
$\tX$ and $\tP$) of our variant of Krawtchouk oscillator.

From \reff{55}, \reff{56}, and \reff{61} we have
\begin{multline}
\qquad V^{-1}A^{-}V=\sqrt{p(1-p)}\left[\left((1-p)N+\frac{\xi}{h}\right)
e^{h\partial_{\xi}}-\right.\\[.3cm]
\left.-\left(pN+\frac{\xi}{h}\right)e^{-h\partial_{\xi}}+
\left((2p-1)N+\frac{2\xi}{h}\right)\right]
;\quad\label{68}\end{multline}
\begin{multline}
\qquad V^{-1}A^{+}V=(1-p)\sqrt{\frac{1-p}{p}}\left(pn+\frac{\xi}{h}\right)
e^{-h\partial_{\xi}}- \\[.3cm]
-p\sqrt{\frac{p}{1-p}}\left((1-p)N-\frac{\xi}{h}\right)e^{h\partial_{\xi}}+
\left((2p-1)N+\frac{2\xi}{h}\right);\label{69}
\end{multline}
\begin{multline}
\qquad V^{-1}H_{AS}V=2p(1-p)N+{\half}+(1-2p)\frac{\xi}{h}-\\[.3cm]
-\left[p\left((1-p)N-\frac{2\xi}{h}\right)e^{h\partial_{\xi}}+
(1-p)\left(pN+\frac{2\xi}{h}\right)
e^{-h\partial_{\xi}}\right];\label{70}
\end{multline}
\begin{equation}\label{71}
V^{-1}A_{0}V=V^{-1}H_{AS}V-\half (N+1).
\end{equation}

Then
\begin{align}
W^{-1}\left(V^{-1}A^{-}V\right)W&=
\sqrt{p(1-p)}\left[(1-x)e^{\partial_x}-xe^{-\partial_x}+2(x-N)\right];
\label{72}\\
W^{-1}\left(V^{-1}A^{-}V\right)W&=
\sqrt{p(1-p)}\left[\frac{1-p}{p}xe^{-\partial_x}-
\frac{p}{1-p}(1-x)e^{\partial_x}+(x-N)\right];
\label{73}\\
W^{-1}\left(V^{-1}H_{AS}^{K}V\right)W&=\!
pN\!+\!\frac{x}{2}\!-\!2px\!-\!\left(p(1-x)e^{\partial_x}\!+
\!(1-p)xe^{-\partial_x}\right);
\label{74}\\
W^{-1}\left(V^{-1}A^0V\right)W&=
W^{-1}\left(V^{-1}H_{AS}^{K}V\right)W-\half(N+1),
\label{75}
\end{align}
and
\begin{align}
\!\!T^{-1}A^{-}T&=\sqrt{p(1-p)}\left((1-x)e^{\partial_x}-
xe^{-\partial_x}+2(x-N)\right); \label{76}\\
\!\!T^{-1}A^{+}T&=\sqrt{p(1-p)}\left(\frac{1-p}{p}xe^{-\partial_x}-
\frac{p}{1-p}(1-x)e^{\partial_x}+2(x-N)\right); \label{77}\\
\!\!T^{-1}H_{AS}^{K}T&=pN+\half+x-2px-
\left(p(1-x)e^{\partial_x}+(1-p)xe^{-\partial_x}\right); \label{78}\\
\!\!T^{-1}A^0T&=T^{-1}H_{AS}^{K}T-\half(N+1). \label{79}
\end{align}

Because the selfadjoint operators
 $\tH$ and $\wt{H}_{AS}=T^{-1}H_{AS}T,$ acting
in Hilbert space
$\HS,$ have the same set of eigenfunctions
$\left\{\tKn\right\}_{n=0}^{N}$
with corresponding eigenvalues
\begin{equation}\label{80}
\lambda_n^{(1)}=\lambda_n(\tH_K)=n+\half,\quad
\lambda_n^{(2)}=\lambda_n(\wt{H}_{AS})=N(n+\half)-n^2,\quad
n=0,1,\ldots,N ,
\end{equation}
it follows that these Hamiltonians are connected by an interesting
relation
\begin{equation}\label{81}
\!\!\tH=-(\wt{H}_{AS}-\half\one)^2+N\wt{H}_{AS}=
-\left(\tH_{AS}-\frac{N+1}{2}\one\right)^2+\frac{N(N+2)}{4}\one
\end{equation}
or
$$
\left(\tH-\frac{N+1}{2}\one\right)+
\left(\tH_{AS}-\frac{N+1}{2}\one\right)^2
=\frac{N^2-2}{4}\one.
$$

\section {Coherent states for generalized oscillator in finite -
dimensional Hilbert space}

\bigskip
\underline{\textbf 1.} In the present section we define coherent
states for generalized oscillator in finite-dimensional space \hmuN
which can be obtained by the truncating procedure described above
from appropriate generalized oscillator in infinite dimensional
space \hmu. In a limiting case when dimension $N$ of the space \hmuN
goes to infinity these coherent states become the coherent states
for the generalized oscillator in \hmu.

We define coherent states in \hmuN by the relation
\begin{equation}\label{82}
\ket{z}=
\Szx{n}{N}\frac{\left(z\ta^{+}_{N}-\bar{z}\ta^{-}_{N}\right)^n}{n!}\ket{0}=
\Szx{n}{N}\left(\sqrt{2}b_{n-1}\right)!C_n(|z|)z^N\ket{n},
\end{equation}
where
\begin{equation}\label{83}
C_n(|z|)=\Szi{m}\gamma_{n, m}\frac{\left(-|z|^2\right)^m}{(n+2m)!}
\end{equation}
Multipliers $\gamma_{n, m}$ are defined from equality
\begin{equation}\label{84}
\gamma_{n, m}=\Szx{k_1}{n}2b_{k_1}^{\,\,2} \Szx{k_2}{k_1+1}2b_{k_2}^{\,\,2}
\ldots\Szx{k_m}{k_{m-1}+1},
\end{equation}
where $b_{k}$ - coefficients from recurrent relations
\reff{18} under an additional condition, that $b_{m}=0$ at $m\geq N.$

Multipliers $\gamma_{n, m}$ satisfy to recurrent relations
\begin{multline}\label{85}
\qquad\gamma_{n+1,m}-\vartheta_{n+1}\gamma_{n,m}=
2b_{n+1}^{\,\,2}\vartheta_{n+2} \gamma_{n+2,m-1};\\
\gamma_{n,0}=1,\quad\gamma_{0,m}=2b_{0}^{\,\,2} \gamma_{1,m-1},
\qquad\qquad
\end{multline}
where
\begin{equation}\label{86}
\vartheta_{n}=\left\{\begin{aligned}
1&\quad\text{if}\quad n\leq N\\
0&\quad\text{if}\quad n>N\\
\end{aligned}\right. .
\end{equation}

Coherent states \reff{82} can be written down in the form
\begin{equation}\label{87}
\ket{z}=\Szx{l}{N}\Syx{n}{l}{\infty}d_{n,l}^{N}
\frac{\left(\sqrt{2}b_{l-1}\right)!}{n!}
(-\bar{z})^{\half(n-l)}z^{\half(n+l)}\ket{l},
\end{equation}
which close to the expression for coherent states of standard boson
FD-oscillator, considered in \cite{17}-\cite{18}.
Coefficients $d_{n, l}^{N}$ satisfy the recurrent relations
\begin{equation}\label{88}
d_{n,l}^{N}=\vartheta_{l}d_{n-1,l-1}^{N}+
2b_{l}^{\,\,2}\vartheta_{l+1}d_{n-1,l+1}^{N},
\end{equation}
with boundary conditions
\begin{equation}\label{89}
d_{n-1,-1}^{N}=0,\qquad d_{0,0}^{N}=1, \qquad d_{n,n+k}^{N}=0,
\quad\text{при}\quad k>0.
\end{equation}
It is easy to prove (compare \cite{17}-\cite{18}), that the solution of
relations \reff{88}, \reff{89} looks like
\begin{equation}\label{90}
d_{n,l}^{N}=\frac{C_N}{\left(2b_{l}^{\,\,2}\right)!}
\Szx{l}{N}\frac{\wt{\psi}_l(x_k)}{\left(\wt{\psi}_N(x_k)\right)^2}x_k^{\,n} ,
\end{equation}
where
\begin{equation}\label{91}
\wt{\psi}_l(\sqrt{2}x)=(\sqrt{2}b_{l-1})!\psi_l^{(0)}(x)
\end{equation}
and $x_k$ are roots of the equation
\begin{equation}\label{92}
\wt{\psi}_{N+1}(x_k)=0,\quad k=0,1,\ldots,N.
\end{equation}
Let us remind that recurrent relations \reff{20} for polynomials
$\psi_n^{(0)}(x)$ differ from relations \reff{18} for
polinomials $\psi_n(x)$ by absence of diagonal members, i.e. $a_n=0.$
The constant $C_N$ in equality \reff{90} can be determined from a
normalization condition. Recurrent relations for polynomials
$\wt{\psi}_{n}(x)$ look like
\begin{equation}\label{91a}
x\wt{\psi}_{n}(x)=\wt{\psi}_{n+1}(x)+2b_{n-1}^{\,\,2}\wt{\psi}_{n-1}(x).
\end{equation}

\bigskip
\underline{\textbf 2.} Now we calculate a value of the constant
$C_N$ in the case of generalized oscillator. (This expression 
reproduces the value of $C_N$ for standard boson FD-oscillator, 
given in \cite{17}-\cite{18} without the proof). To this end we 
rewrite the formula \reff{87} as
\begin{equation}\label{93}
\ket{z}=\Szx{l}{N}C^{(N)}_l\ket{l} ,
\end{equation}
where
\begin{equation}\label{94}
C^{(N)}_l=\Syx{n}{l}{\infty}\frac{(\sqrt{2}b_{l-1})!}{n!}d_{n,l}^{N}
(-\bar{z})^{\half(n-l)}z^{\half(n+l)} ,
\end{equation}
and coefficients $d_{n, l}^{N}$ are determined by the relation \reff{90}.
Then the normalization condition from which we determine $C_N$ takes the
form
\begin{equation}\label{95}
\Szx{l}{N}\left|C^{(N)}_l\right|^2=1.
\end{equation}
In view of \reff{90} and the last condition from \reff{89} the
relation \reff{94} can be rewritten as
\begin{equation}\label{96}
C^{(N)}_l=\frac{C_N}{(\sqrt{2}b_{l-1})!}
\left(-\frac{iz}{|z|}\right)^l\Szx{k}{N}
\frac{\wt{\psi}_l(x_k)}{\left(\wt{\psi}_N(x_k)\right)^2}e^{i|z|x_k}.
\end{equation}
Then
\begin{equation}\label{97}
\left|C^{(N)}_l\right|^2=\frac{C_N^{\,\,2}}{(2b_{l-1}^{\,\,2})!}
\left|A_{l,N}(|z|)\right|^2,
\end{equation}
where
\begin{equation}\label{98}
A_{l,N}(|z|)=\Szx{k}{N}
\frac{\wt{\psi}_l(x_k)}{\left(\wt{\psi}_N(x_k)\right)^2}e^{i|z|x_k}.
\end{equation}
Let's check the validity of the relation
\begin{equation}\label{99}
\frac{\text{d}}{\text{d}|z|}\left|A_{l,N}(|z|)\right|^2=0.
\end{equation}
From \reff{98} we have
\begin{align}
\frac{\text{d}}{\text{d}|z|}\left|A_{l,N}(|z|)\right|^2=
B_{l,N}(|z|)\overline{A_{l,N}(|z|)}+\overline{B_{l,N}(|z|)}A_{l,N}(|z|),
\label{100}\\
B_{l,N}(|z|)=\left(A_{l,N}(|z|)\right)'_{|z|}=\Szx{k}{N}
\frac{\wt{\psi}_l(x_k)}{\left(\wt{\psi}_N(x_k)\right)^2}\,ix_k\,e^{i|z|x_k}.
\label{101}
\end{align}
Because the summands in  the RHS of \reff{100} are complex conjugate to 
each other, we have
\begin{equation}\label{102}
\frac{\text{d}}{\text{d}|z|}
\left(\Szx{l}{N}
\frac{\left|A_{l,N}(|z|)\right|^2}{\left(2b_{l}^{\,\,2}\right)!}\right)=
\left(\Szx{l}{N}\frac{2\text{Re}\left[A_{l,N}(|z|)\overline{B_{l,N}(|z|)}
\right]}{\left(2b_{l}^{\,\,2}\right)!}\right).
\end{equation}
From \reff{91} and recurrent relations \reff{20} it follows                        that
zeros $x_0,x_1,\ldots,x_N$ of the polynomial \reff{92} are located
symmetrically concerning the beginning of coordinates. Then for $l=2p$
$
\wt{\psi}_l(x_k)=\wt{\psi}_l(x_{N-k})
$
and coefficients at
$e^{i|\alpha|x_k}$
and
$e^{-i|\alpha|x_k}$
in $A_{l,N}(|z|)$ are equal. For the case $l=2p+1$ we have
$
\wt{\psi}_l(x_k)=-\wt{\psi}_l(x_{N-k})
$
and coefficients at
$e^{i|\alpha|x_k}$
and at
$e^{-i|\alpha|x_k}$ in $A_{l,N}(|z|)$ have equals module
and are opposite on the sign.
Using notation
\begin{align}
a_{l,k}^{(N)}&=\frac{\wt{\psi}_l(x_k)}{\wt{\psi}_N^{\,\,2}(x_k)},
& b_{l,k}^{(N)}&=a_{l,k}^{(N)} x_k; \label{103} \\
A_{l,N}&=\Szx{k}{N}a_{l,k}^{(N)}e^{i|z|x_k},
& B_{l,N}&=i\Szx{k}{N}b_{l,k}^{(N)}e^{i|z|x_k}, \label{104}
\end{align}
and taking into account above mentioned reasoning we obtain
\begin{equation}\label{105}
\!\!A_{l,N}\!=\!\left\{
\begin{aligned}
2\Szx{k}{m-1}a_{2p,k}^{(N)}\cos(|z|x_k)+a_{2p,m}^{(N)},
&\quad N=2m,\, l=2p;\\
2\Szx{k}{m}a_{2p,k}^{(N)}\cos(|z|x_k),\hphantom{+a_{2p,m}^{(N)}}
&\quad N=2m+1,\, l=2p;\\
\!2i\Szx{k}{m-1}a_{2p+1,k}^{(N)}\sin(|z|x_k)\!+\!a_{2p+1,m}^{(N)},
&\quad N=2m,\, l=2p+1;\\
\!2i\Szx{k}{m}a_{2p+1,k}^{(N)}\sin(|z|x_k),\hphantom{\!+\!a_{2p+1,m}^{(N)}}
&\quad N=2m+1,\, l=2p+1;
\end{aligned}
\right.
\end{equation}
\begin{equation}\label{106}
\!\!\!\!\overline{B_{l,N}}\!=\!\left\{
\begin{aligned}
\!-2\Szx{k}{m-1}b_{2p,k}^{(N)}\sin(|z|x_k)+b_{2p,m}^{(N)},
&\quad N=2m,\, l=2p;\\
\!-2\Szx{k}{m}b_{2p,k}^{(N)}\sin(|z|x_k),\hphantom{+b_{2p,m}^{(N)}}
&\quad N=2m+1,\, l=2p;\\
\!-2i\Szx{k}{m-1}b_{2p+1,k}^{(N)}\cos(|z|x_k)\!+\!b_{2p+1,m}^{(N)},
&\quad N=2m,\, l=2p+1;\\
\!-2i\Szx{k}{m}b_{2p+1,k}^{(N)}\cos(|z|x_k),\hphantom{\!+\!b_{2p+1,m}^{(N)}}
&\quad N=2m+1,\, l=2p+1;
\end{aligned}
\right.
\end{equation}
Note that for obtaining \reff{105} and \reff{106}, we used relations
\begin{gather}\label{107}
a_{2p,k}^{(N)}=a_{2p,N-k}^{(N)},\qquad
a_{2p+1,k}^{(N)}=-a_{2p+1,N-k}^{(N)}, \nonumber \\
b_{2p,k}^{(N)}=-b_{2p,N-k}^{(N)},\qquad
b_{2p+1,k}^{(N)}=b_{2p+1,N-k}^{(N)},\\
b_{l,m}^{2m)}=0, \nonumber
\end{gather}
following from definitions \reff{103} and \reff{104}.
From \reff{105}, \reff{106} and \reff{107} we receive
$$
\text{Re}\left(A_{l,N}(|z|)\overline{B_{l,N}(|z|)}\right)=
$$
\begin{equation}\label{108}
=\left\{
\begin{aligned}
-4\Szx{s,k}{m-1}
a_{2p,s}^{(N)}b_{2p,k}^{(N)}
\cos(|z|x_s)\sin(|z|x_k)-\quad &{}\\
\qquad-2\Szx{k}{m-1}a_{2p,m}^{(N)}b_{2p,k}^{(N)}\sin(|z|x_k),
&\quad N=2m,\, l=2p;\\
-4\Szx{s,k}{m}
a_{2p,s}^{(N)}b_{2p,k}^{(N)}\cos(|z|x_s)\sin(|z|x_k),
&\quad N=2m+1,\, l=2p;\\
4\Szx{s,k}{m-1}
a_{2p+1,s}^{(N)}b_{2p+1,k}^{(N)}
\cos(|z|x_k)\sin(|z|x_s)
&\quad N=2m,\, l=2p+1;\\
4\Szx{s,k}{m}
a_{2p+1,s}^{(N)}b_{2p+1,k}^{(N)}\cos(|z|x_k)\sin(|z|x_s),
&\quad N=2m+1,\, l=2p;
\end{aligned}
\right.
\end{equation}
Substituting \reff{108} in the right-hand side of \reff{102} we see, that
it is sufficient for validity of equality \reff{99} that the coefficients 
at functions  
$\cos(|z|x_s)\sin(|z|x_k)$ ($k,s=0,1,\ldots,(m-1),$
if $N=2m$ and $k,s=0,1,\ldots,m,$ if $n=2m+1,$
$x_s\neq0,$ $x_k\neq0,$)
vanish, i.e.
\begin{equation}\label{*}
\Szx{l}{N}(1-2\delta_{s,k})^l
\frac{\wt{\psi}_l(x_s)\wt{\psi}_l(x_k)}{\left(2b_{l}^{\,\,2}\right)!} x_k=0,
\end{equation}
and also the coefficients at functions $\sin(|z|x_k)$ ($k=0,1,\ldots,(m-1),$
if $N=2m$, $l=2p,$ $x_k\neq0,$) vanish,  i.e.
\begin{equation}\label{**}
\Szx{p}{m}
\frac{\wt{\psi}_{2p}(0)\wt{\psi}_{2p}(x_k)}
{\left(2b_{l}^{\,\,2}\right)!} x_k=0,
\end{equation}

Now we prove the relation \reff{*}. For $0\neq x_k\neq x_s\neq 0$ the
relation \reff{*} is equivalent to equality
\begin{equation}\label{***}
L=x_sx_k\Szx{l}{N}\frac{\wt{\psi}_{l}(x_s)\wt{\psi}_{l}(x_k)}
{\left(2b_{l-1}^{\,\,2}\right)!}=0,
\end{equation}
For the proof of this equality we apply to
$x_s\wt{\psi}_{l}(x_s)$
the recurrent relation \reff{91a}. The result is 
$$
L=x_k\left(\Szx{l}{N-1}
\frac{\wt{\psi}_{l+1}(x_s)
\wt{\psi}_{l}(x_k)}{\left(2b_{l-1}^{\,\,2}\right)!}
+\Syx{l}{1}{N}
\frac{\wt{\psi}_{l-1}(x_s)
\wt{\psi}_{l}(x_k)}{\left(2b_{l-2}^{\,\,2}\right)!}
\right)
$$
If we replace $l$ by $(l+1)$ in the second sum we get
$$
L=x_k\Szx{l}{N-1}\frac{\wt{\psi}_{l+1}(x_s)\wt{\psi}_{l}(x_k)+
\wt{\psi}_{l}(x_s)\wt{\psi}_{l+1}(x_k)}{\left(2b_{l-1}^{\,\,2}\right)!}
=x_k\Pi_N.
$$
Similarly, applying the recurrent relation \reff{91a} to
$x_k\wt{\psi}_{l}(x_k),$ we obtain the equality
$$
L=x_k\Pi_N=x_s\Pi_N.
$$
Because $0\neq x_k\neq x_s\neq 0,$ this relation means $\Pi_N=0$ and
therefore equality \reff{***} is valid.

Now we consider the case $x_k=x_s\neq 0.$ We rewrite \reff{*} as
\begin{equation}\label{****}
\Szx{l}{N}(-1)^l\frac{\wt{\psi}_{l}^{\,2}(x_k)}
{\left(2b_{l-1}^{\,\,2}\right)!}x_k=0.
\end{equation}
Using the recurrent relation \reff{91a}, by induction we obtain
\begin{equation}\label{aa}
x \Szx{l}{N}(-1)^l\frac{\wt{\psi}_{l}(x)}{\left(2b_{l-1}^{\,\,2}\right)!}=
(-1)^N\frac{\wt{\psi}_{N+1}(x)
\wt{\psi}_{N}(x)}{\left(2b_{N-1}^{\,\,2}\right)!},
\end{equation}
from which it follows (taking into account, that $x_k$ are roots 
of function $\wt{\psi}_{N+1}(x)$) the relation \reff{****}.

Now we prove \reff{**}. It is easily to check that
\begin{equation}\label{b}
\wt{\psi}_{2p}(0)=(-1)^p\left(2b_{2p}^{\,\,2}\right)!!.
\end{equation}
Substituting this expression in \reff{**} we see that \reff{**}
is equivalent to relation
\begin{equation}\label{bb}
\Szx{p}{m}(-1)^p
\frac{\wt{\psi}_{2p}(x_k)}{\left(2b_{2p-1}^{\,\,2}\right)!!}=0,
\end{equation}
where $\wt{\psi}_{2m+1}(x_k)=0$  and $x_k\neq 0.$
Using \reff{91a} by induction it is possible to prove identity
\begin{equation}\label{bbb}
x\Szx{p}{m}(-1)^p
\frac{\wt{\psi}_{2p}(x)}{\left(2b_{2p-1}^{\,\,2}\right)!!}=
(-1)^m\frac{\wt{\psi}_{2m+1}(x)}{\left(2b_{2m-1}^{\,\,2}\right)!!},
\end{equation}
from which at $x=x_k$ the relation \reff{bb} follows.

From \reff{99} it follows, that the normalization condition \reff{95}
can be considered for $|z|=0.$ Because from \reff{89}, \reff{90}
and \reff{98} it follows that
\begin{equation}\label{108a}
\left|A_{l,N}(0)\right|=
\left|\Szx{k}{N}\frac{\wt{\psi}'_{l}(x_k)}{\wt{\psi}^2_{N}(x_k)}\right|=
\frac{2b_{l-1}^{\,\,2}}{C_{N}}\left|d_{0,l}^{(N)}\right|=0
\quad\text{при}\quad l>0,
\end{equation}
we obtain
\begin{equation}\label{109}
A_{0,N}(0)=C_{N}\Szx{k}{N}\frac{1}{\wt{\psi}^2_{N}(x_k)}.
\end{equation}

Further, from \reff{97}, \reff{108a} and \reff{109} we have
\begin{equation}\label{110}
\left|C_l^{(N)}(0)\right|=0,\quad\text{при}\quad l>0:
\left|C_0^{(N)}\right|=C_N\Szx{k}{N}\frac{1}{\wt{\psi}^2_{N}(x_k)}.
\end{equation}
Substituting the relation \reff{110} in the normalization condition
\reff{105} we obtain
\begin{equation}\label{111}
C_N=\left(\Szx{k}{N}\left[\wt{\psi}^2_{N}(x_k)\right]^{-2}\right)^{-1}.
\end{equation}
This relation is a generalization of the appropriate formula for a
normalizing constant from \cite{18} which is given there without the proof.
Therefore we shall give bellow the proof of the expression for
this constant also for the case of Hermite polynomials
$\text{He}_{n}(x)$ considered in \cite{18}.

\bigskip
\underline{\textbf 3.} Let us calculate expression in the right-hand side 
of \reff{111} in a case of polynomials 
$
\wt{\psi}_{n}(x)=\text{He}_{n}(x),
$
which fulfill recurrent relations
\begin{align}
\text{He}_{n+1}(x)&=x\text{He}_{n}(x)-n\text{He}_{n-1}(x),\quad
\text{He}_{0}(x)=1;
\label{112} \\
\left(\text{He}_{n+1}(x)\right)'_x&=(n+1)\text{He}_{n}(x)
\label{113}
\end{align}
We consider Lagrange interpolation polynomial  $P_N (x)$ for a
polynomial $\text {He}_{n}(x)$ for interpolation points
$
x_0^{N+1},x_1^{N+1},\ldots,x_N^{N+1}
$
which are roots of a polynomial $\text{He}_{n+1}(x),$ i.e.
\begin{equation}\label{114}
\text{He}_{n+1}(x_k^{N+1})=0,\quad k=0,1,\ldots,N,
\end{equation}
and the values of which in these points are equal to
\begin{equation}\label{115}
P_N(x_k^{N+1})=\left[\text{He}_{N}(x_k^{N+1})\right]^{-1},
\quad k=0,1,\ldots,N.
\end{equation}
It is known \cite{19} that
\begin{align}
P_N(x)&=\Szx{k}{N}\frac{\text{He}_{N+1}(x)}
{(x-x_k^{N+1})\text{He}_{N}(x_k^{N+1})\text{He}'_{N+1}(x_k^{N+1})}=
\nonumber\\
&=\frac{1}{N+1}\Szx{k}{N}\frac{\text{He}_{N+1}(x)}
{(x-x_k^{N+1})\text{He}^2_{N}(x_k^{N+1})} .\label{116}
\end{align}
From \reff{116} and \reff{111} we see that $C_N^{-1}$ coincides
with the leading coefficient of the polynomial $(N+1)P_N(x).$
We introduce an auxiliary polynomial
\begin{equation}\label{117}
\Phi_{2N}(x)=P_N(x)\text{He}_{N}(x)-1,
\end{equation}
leading coefficient of which  is equal to
\begin{equation}\label{118}
K_N=\frac{1}{(N+1)C_N}.
\end{equation}
From \reff{115} and \reff{117} it follows that
$\Phi_{2N}(x_k^{N+1})=0,$ \quad $k=0,1,\ldots,N,$
so the polynomial
$\Phi_{2N}(x)$ can be divided on the polynomial $\text{He}_{N+1}(x)$
\begin{equation}\label{119}
\Phi_{2N}(x)=\text{He}_{N+1}(x)Q_{N-1}(x),
\end{equation}
where $Q_{N-1}(x)$ is a polynomial of the order $N-1.$
Because from \reff{112} it follows that the leading coefficient 
of a polynomial $\text{He}_{N}(x)$ is equal to 1, we see from 
\reff{118} and \reff{119} that the leading coefficients of a 
polynomial $Q_{N-1} (x)$ is equal to $K_N.$
Further, from \reff{117} and \reff{119} we have
\begin{equation}\label{120}
\text{He}_{N+1}(x_k^{N}) Q_{N-1}(x_k^{N})=-1,\quad k=0,1,\ldots,N-1,
\end{equation}
where $x_k^{N}$ are roots of the polynomial
$\text{He}_{N}(x).$ Taking into account \reff{112}, we have
\begin{equation}\label{121}
\text{He}_{N+1}(x_k^{N})=-N\text{He}_{N-1}(x_k^{N}),\quad k=0,1,\ldots,N-1,
\end{equation}
Substituting \reff{121} in \reff{120} and taking into account \reff{115},
we obtain
\begin{equation}\label{122}
Q_{N-1}(x_k^{N})=\frac{1}{N\text{He}_{N-1}(x_k^{N})}=
\frac{1}{P_{N-1}(x_k^N)},\quad k=0,1,\ldots,N-1.
\end{equation}
Because two polynomials $Q_{N-1}(x)$ and $P_{N-1}(x),$ coinciding
in $N$ points $x_0^{N},x_1^{N},\ldots,x_{N-1}^{N},$ are equal,
the leading coefficients of a polynomial $P_{N-1}(x)$ is equal to
\begin{equation}\label{123}
N\,K_N=\frac{N}{(N+1)C_N}.
\end{equation}
Taking into account that $P_{0}(x)=1$ and continuing by induction, 
we received
\begin{equation}\label{124}
C_N=\frac{N!}{N+1}.
\end{equation}

\bigskip
Returning to the general case, we receive, taking into account
\reff{93}, \reff{96} and \reff{111}, an analytical expression 
for coherent states for the generalized oscillator
\begin{equation}\label{125}
\ket{z}=\left(\Szx{k}{N}\left[\wt{\psi}_{N}(x_k)\right]^{-2}\right)^{-1}
\Szx{l}{N}\frac{\left(-i\frac{z}{|z|}\right)^l}{\left(\sqrt{2}b_{l-i}\right)!}
\left(\Szx{k}{N}\frac{\wt{\psi}_{l}(x_k)}{\wt{\psi}_{N}^2(x_k)}
e^{i|z|x_k}\right)\ket{l},
\end{equation}
where $x_k$ are roots of the equation \reff{92}.

\bigskip
\underline{\textbf 4.} The completeness for coherent
states \reff{125} can be proved in the standard way. For
To construct a measure
\begin{equation}\label{126}
\text{d}\mu(|z|^2)=\wt{W}(|z|^2)\text{d}^2z
\end{equation}
participating in  the decomposition of unit
\begin{equation}\label{127}
\iint_{\BC}\wt{W}(|z|^2)\ket{z}\bra{z}\text{d}^2z=\one ,
\end{equation}
where $\text{d}^2z=\text{dRe}z\text{dIm}z,$ we designate $x=|z|^2$ 
and substitute \reff{125} in \reff{127}. Using \reff{98} and 
\reff{111}, we obtain
\begin{equation}\label{128}
\Szx{l}{N}\frac{\pi C_N^2}{(2b_{l-1}^{\,\,2})!}
\int_0^{\infty}\wt{W}(x)\left|A_{l,N}(\sqrt{x})\right|^2\text{d}x
\ket{l}\bra{l}=1.
\end{equation}
Using the notation
\begin{equation}\label{129}
W(x)=\pi\wt{W}(x)
\end{equation}
we obtain the following condition
\begin{equation}\label{130}
\int_0^{\infty}W(x)|A_{l,N}|^2\text{d}x=(2b_{l-1}^{\,\,2})!,\quad
l=0,1,\ldots,N
\end{equation}
for definition of a function $W(x)$ 
We shall not discuss here the solution of this (trigonometrical)
moments problem \cite{20}.

\bigskip
\underline{\textbf 5. \,} Coherent states for Krawtchouk oscillator
can be obtained from the general formula \reff{125} if we 
replace functions $\wt{\psi}_{n} (x)$ by the Krawtchouk polynomials 
with recurrent relations
\begin{equation}\label{131}
x\wt{\psi}_{n}(x)=
\wt{\psi}_{n+1}(x)+2p(1-p)n(N-n+1)\wt{\psi}_{n-1}(x),\quad
\wt{\psi}_{0}(x)=1,
\end{equation}
Note, that coherent states obtained in this way
differ from "spin" coherent states for Krawtchouk oscillator
given in \cite{14} as well as from "phase" coherent states defined in the
work \cite{21}.

\bigskip
\bigskip

Department of Mathematics,

St.Petersburg University of Telecommunications,

191065, Moika  61, St.Petersburg, Russia

{vadim@VB6384.spb.edu}

  \medskip

Department of Mathematics,

University of Defense Technical Engineering,

191123, Zacharievskaya 22, St.Petersburg, Russia

{evd@pdmi.ras.ru}

\end {document}